\documentclass[conference]{IEEEtran}
\usepackage[left=0.625in, right=0.625in, top=0.75in,bottom=.95in]{geometry}
\IEEEoverridecommandlockouts
\usepackage{tabularx}
\usepackage[T1]{fontenc}
\usepackage{algorithm}
\usepackage[noend]{algpseudocode}
\usepackage[dvipdfmx]{graphicx} 
\usepackage{bmpsize}
\usepackage{amsmath,amssymb,amsfonts,stfloats}
\usepackage{array}
\usepackage{textcomp}
\usepackage{paralist}
\usepackage{lettrine}
\usepackage{comment}
\usepackage{cite}
\usepackage[table,xcdraw]{xcolor}
\usepackage{multirow}
\usepackage{cancel}
\usepackage{booktabs}
\usepackage{mathtools}
\usepackage{array}
\usepackage{booktabs}
\usepackage[caption=false,font=footnotesize]{subfig}
\usepackage{url}
\usepackage{acronym}
\usepackage{multicol}
\usepackage{makecell}
\usepackage{cases}
\usepackage{overpic}

\usepackage{xspace}

\newcommand{\rev}[1]{\textcolor{black}{#1}}

\newcommand{\refs}[2]{(\ref{#1})-(\ref{#2})}
\newcommand{\Figref}[1]{Fig.~\ref{#1}}

\newcommand{\Secref}[1]{Section~\ref{#1}}

\newcommand{\bbf}{\mathbf{f}}
\newcommand{\bg}{\mathbf{g}}

\newcommand{\bn}{\mathbf{n}}

\newcommand{\bs}{\mathbf{s}}

\newcommand{\bu}{\mathbf{u}}
\newcommand{\bv}{\mathbf{v}}
\newcommand{\bw}{\mathbf{w}}
\newcommand{\by}{\mathbf{y}}
\newcommand{\bx}{\mathbf{x}}
\newcommand{\bz}{\mathbf{z}}

\newcommand{\bH}{\mathbf{H}}

\newcommand{\bU}{\mathbf{U}}

\newcommand{\bW}{\mathbf{W}}

\newtheorem{remark}{Remark}

\newcommand{\herm}{^{H}}
\newcommand{\trans}{^{T}}

\DeclareMathOperator{\E}{\mathsf{E}}

\newcommand{\EX}[1]{\E\left\{{#1}\right\}}

\newcommand{\snorm}[1]{\left\Vert #1 \right\Vert^2}
\newcommand{\sabs}[1]{{ \left\vert #1 \right\vert^2 }}

\newcommand{\mC}{\mathbb{C}}
\newcommand{\mR}{\mathbb{R}}
\newcommand{\setK}{\mathcal{K}}

\newcommand{\UE}{^{\mathsf{UE}}}
\newcommand{\BS}{^{\mathsf{gNB}}}
\newcommand{\IAB}{^{\mathsf{IAB}}}
\newcommand{\pMaxBS}{p\BS_{\mathsf{max}}}
\newcommand{\pMaxIAB}{p\IAB_{\mathsf{max}}}
\newcommand{\pMaxUE}{p\UE_{\mathsf{max}}}
\newcommand{\wK}{\widetilde{K}}
\newcommand{\SEuIAB}{\mathsf{SE}^{\mathsf{u,IAB}}}
\newcommand{\SEdIAB}{\mathsf{SE}^{\mathsf{d,IAB}}}
\newcommand{\SEuBS}{\mathsf{SE}^{\mathsf{u,gNB}}}
\newcommand{\SEdBS}{\mathsf{SE}^{\mathsf{d,gNB}}}
\newcommand{\SINRuIAB}{\mathsf{SINR}^{\mathsf{u,IAB}}}
\newcommand{\SINRdIAB}{\mathsf{SINR}^{\mathsf{d,IAB}}}
\newcommand{\SINRsIAB}{\mathsf{SINR}^{\mathsf{x,IAB}}}
\newcommand{\SINRuBS}{\mathsf{SINR}^{\mathsf{u,gNB}}}
\newcommand{\SINRdBS}{\mathsf{SINR}^{\mathsf{d,gNB}}}
\newcommand{\INTuIAB}{\mathsf{INT}^{\mathsf{u,IAB}}}
\newcommand{\INTdIAB}{\mathsf{INT}^{\mathsf{d,IAB}}}
\newcommand{\INTuBS}{\mathsf{INT}^{\mathsf{u,gNB}}}
\newcommand{\INTdBS}{\mathsf{INT}^{\mathsf{d,gNB}}}
\newcommand{\SINRxy}{\mathsf{SINR}^{\mathsf{x,y}}}
\newcommand{\SExy}{\mathsf{SE}^{\mathsf{x,y}}}
\newcommand{\uIAB}{^{\mathsf{u,IAB}}}
\newcommand{\dIAB}{^{\mathsf{d,IAB}}}
\newcommand{\sIAB}{^{\mathsf{x,IAB}}}
\newcommand{\uBS}{^{\mathsf{u,gNB}}}
\newcommand{\dBS}{^{\mathsf{d,gNB}}}
\newcommand{\setwK}{{\mathcal{I}}}

\makeatletter
\g@addto@macro\normalsize{%
  \setlength\abovedisplayskip{2.3pt}
  \setlength\belowdisplayskip{2.3pt}
}
\makeatother

\renewcommand{\baselinestretch}{.97}

\begin{document}
\bstctlcite{IEEEexample:BSTcontrol}

\title{Joint Optimization of Uplink and Downlink Power in Full-Duplex Integrated Access and Backhaul}
\author{\IEEEauthorblockN{Giovanni Interdonato\IEEEauthorrefmark{1}, Silvia Mura\IEEEauthorrefmark{2}, Marouan Mizmizi\IEEEauthorrefmark{2}, Stefano Buzzi\IEEEauthorrefmark{1}\IEEEauthorrefmark{2} and Umberto Spagnolini\IEEEauthorrefmark{2}}
\IEEEauthorblockA{\IEEEauthorrefmark{1}\textit{Dept. of Electrical and Information Engineering, University of Cassino and Southern Lazio, 03043 Cassino, Italy.}\\
\IEEEauthorrefmark{2}\textit{Dept. of Electronics, Information and Bioengineering, Politecnico di Milano, 20133 Milan, Italy.}\\
\texttt{\small\{giovanni.interdonato,buzzi\}@unicas.it}\\[-0.5ex]
\texttt{\small\{silvia.mura,marouan.mizmizi,umberto.spagnolini\}@polimi.it}}\\[-7.0ex]
\thanks{Paper supported by the EU under the Italian \textit{National Recovery and Resilience Plan} of NextGenerationEU, partnership on “Telecommunications of the Future” (PE00000001-program RESTART, structural project 6GWINET).}}

\begin{figure*}[t!]
\normalsize
This paper has been accepted for publication in the proceedings of the IEEE International Conference on Communications (ICC) 2025, 8--12 June 2025, Montreal, Canada.

\

\textcopyright~2025 IEEE. Personal use of this material is permitted. 
Permission from IEEE must be obtained for all other uses, in any current or future media, including reprinting/republishing this material for advertising or promotional purposes, creating new collective works, for resale or redistribution to servers or lists, or reuse of any copyrighted component of this work in other works.

\


\vspace{17cm}
\end{figure*}

\maketitle

\vspace{-10mm}
\begin{abstract}
We examine the performance of an Integrated Access and Backhaul (IAB) node as a range extender for beyond-5G networks, focusing on the significant challenges of effective power allocation and beamforming strategies, which are vital for maximizing users' spectral efficiency (SE). We present both max-sum SE and max-min fairness power allocation strategies, to assess their effects on system performance. The results underscore the necessity of power optimization, particularly as the number of users served by the IAB node increases, demonstrating how efficient power allocation enhances service quality in high-load scenarios. The results also show that the typical line-of-sight link between the IAB donor and the IAB node has rank one, posing a limitation on the effective SEs that the IAB node can support.%
\end{abstract}

\section{Introduction}
Integrated Access and Backhaul (IAB) is a key technology for \textit{beyond-5G} systems, enhancing coverage, capacity, and flexibility~\cite{madapatha2020integrated}. Traditional cellular networks rely on fixed backhaul infrastructure, such as fiber or microwave links, to connect cell sites to the core network. However, deploying these solutions, especially in dense urban or remote areas, can be costly and logistically challenging. IAB addresses these issues by using the same wireless spectrum and resources for both access (connecting users) and backhaul (connecting base stations), eliminating the need for separate infrastructure. Wireless backhauling is particularly appealing in the millimeter-wave (mmWave) bands due to the wide bandwidth available~\cite{zhang2021survey}.

With IAB, only a few nodes, called \textit{IAB donors}, require fiber-based backhaul connections, while the remaining \textit{IAB nodes} function as relays, wirelessly forwarding backhaul traffic for multi-hop communications~\cite{Dhillon2015,Polese2020}.  
IAB nodes have baseband capabilities and perform transceiver signal processing, including transmit precoding, receive combining, and channel estimation. 
IAB saw limited deployment in \textit{Long-Term Evolution} due to high spectrum costs, single-hop backhauling, and inflexible topologies. Renewed interest in Release 16 aimed to enable flexible, effective integration into 5G networks, and today, Release 18 is developing enhanced \textit{New Radio} backhauling~\cite{3GPPTS38174}. Key improvements include simultaneous multi-node communication via spatial division multiplexing, boosting robustness, spectral efficiency (SE), and performance. Full-duplex (FD) operation~\cite{Zhang2021IAB} further enhances SE and reduces latency, allowing concurrent transmission and reception without time- or frequency-division multiplexing.

The main challenge in FD-IAB networks is mitigating self-interference, which limits SE.  
The authors in~\cite{Zhang2024IAB} analyze in-band-FD wideband IAB performance using stochastic geometry and derive closed-form signal-to-interference-plus-noise ratio (SINR) coverage expressions. In~\cite{lagunas2017power}, a heuristic power allocation strategy enhances the downlink sum rate under backhaul capacity constraints. A joint optimization of power, time splitting, user association, and beamforming was proposed in~\cite{kwon2019joint} to maximize weighted sum rate in mmWave IAB networks. The benefits of combining \textit{small cells} and massive multiple-input multiple-output (MIMO) in in-band wireless backhaul were analyzed in~\cite{li2015small, chen2019user}, while~\cite{vu2019joint} studied path selection and rate allocation for multi-hop self-backhaul mmWave networks. The sum-rate maximization problem in cell-free massive MIMO was addressed in~\cite{jazi2023integrated}.

\textit{Contribution}: We propose a joint uplink and downlink power optimization framework for in-band FD mmWave IAB networks, aiming to maximize either the overall sum SE or the sum of the minimum SEs. \rev{While most IAB literature defines \textit{full duplex} in the context of the IAB node’s operation, we extend this notion to network-wide simultaneous uplink-downlink operation, where the IAB donor and IAB node function in opposite transceiver modes---an in-band network-enabled distributed FD implementation~\cite{Thomsen2016}.}
Our optimization approach addresses both backhaul capacity constraints and \rev{cross-link interference from simultaneous uplink and downlink transmissions.}  
Unlike prior works, we globally solve the joint uplink-downlink power optimization problem, considering the achievable SE of access links at both the IAB donor and the IAB node, while dynamically adjusting backhaul link capacity as needed. \rev{In contrast,~\cite{chen2019user} focuses only on downlink transmissions, assuming FD operation at the IAB nodes.}

\section{System Model} \label{Sec:SystemModel}
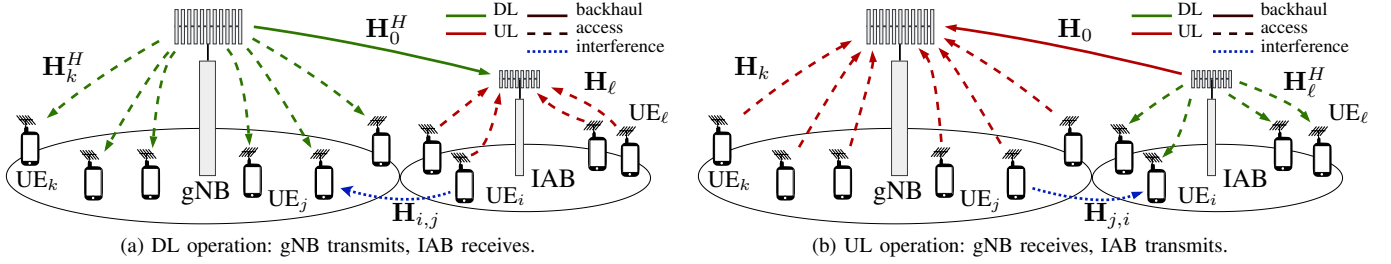
\begin{figure*}[!t]
    \centering
    \vspace{1mm}%
    \subfloat[DL operation: gNB transmits,  IAB receives.]{%
        \begin{overpic}[abs,unit=1mm,width=.96\columnwidth]{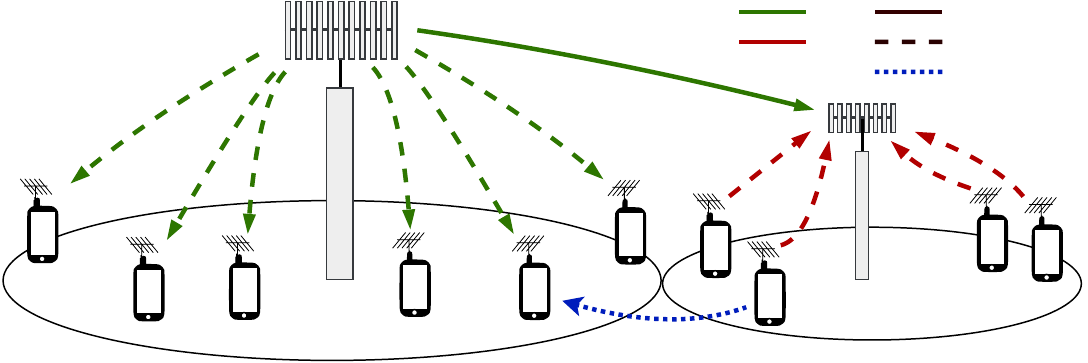}
            \put(23.5,3.5){gNB}
            \put(70,5){IAB}
            \put(65,28){\scriptsize DL}
            \put(65,25){\scriptsize UL}
            \put(76,28){\scriptsize backhaul}
            \put(76,25.5){\scriptsize access}
            \put(76,23){\scriptsize interference}
            \put(5,20){$\bH\herm_k$}
            \put(48,25){$\bH\herm_0$}
            \put(1.5,5){\small UE$_k$}
            \put(35,2.5){\small UE$_j$}
            \put(64,3){\small UE$_i$}
            \put(77,18){$\bH_\ell$}
            \put(83,14){\small UE$_\ell$} 
            \put(51.5,0.5){$\bH_{i,j}$}
            \label{subfig:DL}
        \end{overpic}%
    }\hspace{6mm}%
    \subfloat[UL operation: gNB receives, IAB transmits.]{%
        \begin{overpic}[abs,unit=1mm,width=.96\columnwidth]{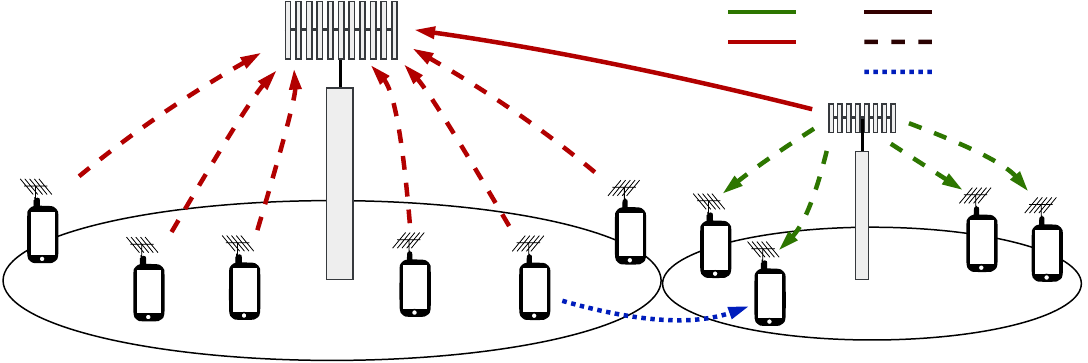}
            \put(23.5,3.5){gNB}
            \put(70,5){IAB}
            \put(64,28){\scriptsize DL}
            \put(64,25){\scriptsize UL}
            \put(75,28){\scriptsize backhaul}
            \put(75,25.5){\scriptsize access}
            \put(75,23){\scriptsize interference}
            \put(5,20){$\bH_k$}
            \put(48,25){$\bH_0$}
            \put(1.5,5){\small UE$_k$}
            \put(35,2.5){\small UE$_j$}
            \put(64,3){\small UE$_i$}
            \put(78,18){$\bH\herm_\ell$}
            \put(84,14){\small UE$_\ell$}
            \put(51.5,0.5){$\bH_{j,i}$}
            \label{subfig:UL}
        \end{overpic}%
    }%
    \caption{An example of an in-band full-duplex IAB network with \Figref{subfig:DL} depicting the gNB DL operation and \Figref{subfig:UL} depicting the gNB UL operation.}
    \vspace*{-4mm}
    \label{fig:RefScen}
\end{figure*}

Let us consider an in-band-FD IAB network with single hop, as shown in the illustrative example of Fig.~\ref{fig:RefScen}. 
We assume that $K$ user equipments (UEs) are served by the gNodeB (gNB), while $\widetilde{K}$ UEs, which are not in the serving coverage area of the gNB, are served by the IAB node. Hence, the gNB also serves as IAB donor, while the IAB node acts as range extender for the gNB.
All the $K\!+\!\widetilde{K}$ UEs are served on the same mmWave time-frequency resource. We denote by $\setK \!=\!\{1,\ldots,K\}$ and $\setwK\!=\!\{K\!+\!1,\ldots,K\!+\!\wK\}$ the set of the indices of the UEs served by the gNB and the IAB node, respectively.
The gNB, IAB node, and UEs are equipping a fully-digital MIMO array with $N\BS$, $N\IAB$, and $N\UE$ antennas, respectively, and as many RF chains. It holds that $N\BS \!\geq\! K\!+\!1$ with $K$ streams dedicated to the served UEs and one to the backhaul link with the IAB node. Similarly, for the IAB node, it holds $N\IAB \!\geq\! \widetilde{K}\!+\!1$. The system operates in time-division duplex (TDD) according to the gNB operating condition. Specifically, we distinguish two stages: $(i)$ \textit{Downlink (DL) operation}, wherein the gNB is in transmitter mode while the IAB node is in receiver mode, as shown in Fig.~\ref{subfig:DL}. Hence, the gNB is serving the $K$ UEs and the IAB node on its DL, while the IAB node is receiving the signals from the served $\widetilde{K}$ UEs. $(ii)$ \textit{Uplink (UL) operation}, wherein the gNB is in receiver mode while the IAB node is in transmitter mode, as shown in Fig.~\ref{subfig:UL}. Hence, the gNB receives the signal from the $K$ UEs and the IAB node on its UL, while the IAB node serves the $\widetilde{K}$ UEs. \rev{At each TDD stage, the roles of gNB and IAB node are reversed, enabling simultaneous UL and DL transmissions.}

Any of the channels follow the cluster-based multipath channel model detailed in \cite{meijerink2014physical,3GPPTR38901}: 
\begin{align}
    \mathbf{H} & = \sqrt{\frac{N_{t}\,N_{r}}{N_c\,N_\ell}} \sum_{c=1}^{N_c} \sum_{\ell=1}^{N_\ell}  \alpha_{c,\ell} \, \mathbf{a}_{N_{r}}(\boldsymbol{\theta}_{c,\ell}) \mathbf{a}^T_{N_t}(\boldsymbol{\phi}_{c,\ell}) \, ,\label{eq:inChannel}
\end{align}
where \rev{$N_t$ and $N_r$ denote the number of antennas at the transmitter and receiver, respectively,} $N_c$ is the number of scattering clusters, $N_\ell$ is the number of paths per cluster (herein assumed to be equal for all the clusters), $\alpha_{c,\ell}\sim \mathcal{CN}\left(0, \Omega_{c,\ell}\right)$ is the scattering amplitude of the $\ell$-th path of the $c$-th cluster, with power $\Omega_{c,\ell}$, $\mathbf{a}_{N_{t}}(\boldsymbol{\phi}_{c,\ell})$ and $\mathbf{a}_{N_{r}}(\boldsymbol{\theta}_{c,\ell})$ are the transmitter and receiver array response vectors, function of the angles of departure $\boldsymbol{\phi}_{c,\ell}$ and arrival $\boldsymbol{\theta}_{c,\ell}$, respectively. 
\rev{The channel model in~\eqref{eq:inChannel} consists of a dominant line-of-sight (LoS) path and $N_c N_{\ell} \!-\!1$ non-LoS (NLoS) paths due to the scattering clusters.}
As for the backhaul link from the gNB to the IAB node, we assume $N_c \!=\! 1, N_\ell \!=\! 1$, hence LOS scenario, due to the antenna high altitude as gNB and IAB node are typically on top of buildings. \rev{We assume the \textit{block-fading model} to simplify the analysis and focus on core principles without the excessive analytical complexities of practical multicarrier schemes.}%

\subsection{gNB Downlink Operation}

The DL signal transmitted by the gNB to the $K$ served UEs and to the IAB node is given by
\begin{align}\label{eq:BS:power-constraint}
\bx\BS \!=\! \sum\limits^K_{k=0}\sqrt{\eta\BS_k} \bbf\BS_k x\BS_k \, \in \mC^{N\BS},
\end{align}
where the index $k\!=\!0$ refers to the IAB node, $\{x\BS_k\}$ are zero-mean uncorrelated data symbols with $\EX{\lvert x\BS_k \rvert^2} \!=\! 1$, and $\{\bbf\BS_k\}$ are the precoding vectors intended for the receiver $k$, with $\lVert \bbf\BS_k \rVert^2 \!=\! 1,~k\!=\!0,\ldots,K$. Moreover, $\{\eta\BS_k\}$ are the DL transmit powers subject to the power constraint
\begin{align}
\EX{\snorm{\bx\BS}} = \sum\nolimits^K_{k=0}\eta\BS_k \leq \pMaxBS \, ,
\end{align}
where $\pMaxBS$ is the maximum power at the gNB.
Similarly, the UL signal transmitted by UE $i$, $i \!\in\! \setwK$, to the IAB node is
\begin{align}
\bx\UE_i = \sqrt{\eta\UE_i}\bbf\UE_i x\UE_i \, \in \mC^{N\UE},
\end{align}
where $\{x\UE_i\}$ are zero-mean uncorrelated data symbols with $\EX{\lvert x\UE_i \rvert^2} \!=\! 1$, and $\{\bbf\UE_i\}$ are the precoding vectors, with $\lVert \bbf\UE_i \rVert^2 \!=\! 1,~i\!\in\! \setwK$. Moreover, $\{\eta\UE_i\}$ are the UL transmit powers subject to the power constraint
\begin{align} \label{eq:UE:power-constraint}
\EX{\snorm{\bx\UE_i}} = \eta\UE_i \leq \pMaxUE\,, 
\end{align}
where $\pMaxUE$ is the maximum transmit power at any UE.

The DL signal received by the $k$-th UE, $k\!\in\!\setK$, under the gNB coverage, is thus given by  
\begin{align}
    \by\UE_k  \!=\! \bH\herm_k \bx\BS \!+\! \sum\limits_{i \in \setwK} \bH_{i,k} \bx\UE_i \!+\! \bn\UE_k \, ,
\end{align}
where the second term represents the multiuser interference due to the simultaneous UL transmissions of the UEs under the IAB node coverage, and $\bn\UE_k$ is the AWGN vector at UE $k$, whose elements have variance $\sigma_{\mathsf{UE}}^2$. Let $\bv\UE_k$ be the combining vector used by UE $k$, $k\!\in\!\setK$, to linearly process $\by\UE_k$ as $r\UE_k \!=\! {(\bv\UE_k)}\herm \by\UE_k$. Then, by letting $\Tilde{\bn}\UE_k \!=\!{(\bv\UE_k)}\herm\bn\UE_k$, we have
\begin{align*} 
    \label{eqref:DL:UE:signal-processed}
    r\UE_k &\!=\! {(\bv\UE_k)}\herm\bH\herm_k \bx\BS \!+\! {(\bv\UE_k)}\herm \sum\limits_{i \in \setwK} \bH_{i,k} \bx\UE_i \!+\! \Tilde{\bn}\UE_k \nonumber \\
    &\!=\! \sum\limits^{K}_{j = 0}\sqrt{\eta\BS_j} \bg_{k,k,j}^{\mathsf{UE\text{-}gNB}} x\BS_j \!+\! \sum\limits_{i \in \setwK} \! \sqrt{\eta\UE_i} \Tilde{\bg}_{k,i}^{\mathsf{UE\text{-}UE}} x\UE_i \!+\! \Tilde{\bn}\UE_k,
\end{align*} 
where $\bg_{k,i,j}^{\mathsf{C\text{-}P}} \!=\! {(\bv^{\mathsf{C}}_k)}\herm\bH\herm_i \bbf^{\mathsf{P}}_j$, with $\mathsf{C} \!=\! \mathsf{P} \!=\! \{\mathsf{UE},\,\mathsf{IAB},\,\mathsf{gNB} \}$ referring to the node wherein the combining, respectively, the precoding takes place. Besides, $\Tilde{\bg}_{k,i}^{\mathsf{UE\text{-}UE}} \!=\! {(\bv^{\mathsf{UE}}_k)}\herm\bH_{i,k} \bbf^{\mathsf{UE}}_i$. 

The signal received at the IAB node sums up the signal transmitted by the gNB on its DL and the superposition of the signals transmitted by the IAB-node UEs on their UL as
\begin{align}
\by\IAB = \bH\herm_0\bx\BS + \sum\limits_{i \in \setwK} \bH_i \bx\UE_i + \bn\IAB \, \in \mC^{N\IAB}\, ,
\end{align}
where $\bn\IAB$ is the AWGN vector at the IAB node, whose elements have variance $\sigma_{\mathsf{IAB}}^2$. The IAB node combines $\by\IAB$ to decode $x\BS_0$, by using the combining vector $\bv\IAB_0$, as $r\IAB_0 \!=\! {(\bv\IAB_0)}\herm\by\IAB$. By letting $\Tilde{\bn}_0\IAB \!=\! {(\bv\IAB_0)}\herm\bn\IAB$, we have 
\begin{align*}
r\IAB_0 &\!=\! {(\bv\IAB_0)}\herm\bH\herm_0\bx\BS \!+\! {(\bv\IAB_0)}\herm \!\sum\nolimits_{i \in \setwK} \bH_i \bx\UE_i \!+\! \Tilde{\bn}_0\IAB \nonumber \\
&\!=\! \sum\limits^K_{k=0}\!\sqrt{\eta\BS_k} \bg_{0,0,k}^{\mathsf{IAB\text{-}gNB}}  x\BS_k \!+\! \sum\limits_{i \in \setwK} \!\sqrt{\eta\UE_i} \hat{\bg}_{0,i,i}^{\mathsf{IAB\text{-}UE}} x\UE_i \!+\! \Tilde{\bn}_0\IAB,
\end{align*}
where $\hat{\bg}_{k,i,j}^{\mathsf{C\text{-}P}} \!=\! {(\bv^{\mathsf{C}}_k)}\herm\bH_i \bbf^{\mathsf{P}}_j$ with $\mathsf{C} \!=\! \mathsf{P} \!=\! \{\mathsf{UE},\,\mathsf{IAB},\,\mathsf{gNB} \}$ referring to the node wherein the combining, respectively, the precoding takes place. Similarly, the IAB node combines the received signal by using $\bv\IAB_i$ to obtain $x\UE_i$, $i \!\in\!\setwK$, as $r\IAB_i \!=\! {(\bv\IAB_i)}\herm\by\IAB$. By letting $\Tilde{\bn}_i\IAB \!=\! {(\bv\IAB_i)}\herm\bn\IAB$, we have
\begin{align*}
r\IAB_i &\!=\! {(\bv\IAB_i)}\herm\bH\herm_0\bx\BS + {(\bv\IAB_i)}\herm\sum\nolimits_{\ell\in\setwK} \bH_{\ell} \bx\UE_{\ell} + \Tilde{\bn}_i\IAB \nonumber \\
&\!=\! \sum\limits_{\ell\in\setwK} \!\sqrt{\eta\UE_{\ell}} \hat{\bg}_{i,\ell,\ell}^{\mathsf{IAB\text{-}UE}} x\UE_{\ell} \!+\! \sum\limits^K_{k=0}\!\sqrt{\eta\BS_k} \bg_{i,0,k}^{\mathsf{IAB\text{-}gNB}} x\BS_k \!+\! \Tilde{\bn}_i\IAB.
\end{align*}

\subsection{gNB UL Operation}
The UL signal transmitted by UE $k$, $k \!\in\!\setK$, to the gNB is
\begin{align}
\bx\UE_k = \sqrt{\eta\UE_k}\bbf\UE_k x\UE_k \, \in \mC^{N\UE}, 
\end{align}
which is subject to the constraint in~\eqref{eq:UE:power-constraint}.
Besides, the IAB node transmits the following signal to the gNB and its UEs 
\begin{align}
\bx\IAB \!=\! \sqrt{\eta\IAB_0} \bbf\IAB_0 x\IAB_0 \!+\! \sum\nolimits_{i\in\setwK} \!\sqrt{\eta\IAB_i} \bbf\IAB_i x\IAB_i \! \in \! \mC^{N\IAB}, 
\end{align}
which is subject to the following power constraint
\begin{align}
\EX{\snorm{\bx\IAB}} = \eta\IAB_0+\sum\nolimits_{i\in\setwK} \eta\IAB_i \leq \pMaxIAB\, , 
\end{align}
upon the assumptions of zero-mean uncorrelated data symbols $\{x\IAB_i\}$ with $\EX{\lvert{x\IAB_i}\rvert^2} \!=\! 1$, and unit-power precoding vector, $\lVert{\bbf\IAB_i}\rVert^2 \!=\! 1, \, i\!\in\!\setwK\cup\{0\}$. 
The signal received at the gNB is
\begin{align}
\!\by\BS \!&= \sum\nolimits_{k\in\setK} \sqrt{\eta\UE_k} \bH_k \bbf\UE_k x\UE_k + \sqrt{\eta\IAB_0} \bH_0 \bbf\IAB_0 x\IAB_0 \nonumber \\
&\quad+ \bH_0 \sum\nolimits_{i \in \setwK} \sqrt{\eta\IAB_i} \bbf\IAB_i x\IAB_i + \bn\BS  \, , 
\end{align}
where $\bn\BS$ is the AWGN vector at the gNB, whose elements have variance $\sigma^2_{\mathsf{gNB}}$.
To decode $x\UE_k$, with $k \!\in\! \setK$, the gNB combines $\by\BS$ by using the combining vector $\bv\BS_k$ as $r\BS_k \!=\! {(\bv\BS_k)}\herm \by\BS$. Then, we have  
\begin{align}
\label{eq:rBS_k}
r\BS_k &=\! \sum\nolimits_{j \in \setK} \sqrt{\eta\UE_j} \hat{\bg}_{k,j,j}^{\mathsf{gNB\text{-}UE}}  x\UE_j \!+\! \sqrt{\eta\IAB_0} \hat{\bg}_{k,0,0}^{\mathsf{gNB\text{-}IAB}} x\IAB_0 \nonumber \\
&\quad  +\! \sum\nolimits_{i \in \setwK} \!\!\sqrt{\eta\IAB_i} \hat{\bg}_{k,0,i}^{\mathsf{gNB\text{-}IAB}} x\IAB_i \!+\! \Tilde{\bn}_k\BS,
\end{align} 
where $\Tilde{\bn}_k\BS \!=\! {(\bv\BS_k)}\herm\bn\BS$.
Similarly, to decode $x\IAB_0$ the gNB combines $\by\BS$ by using the combining vector $\bv\BS_0$ to obtain $r\BS_0 \!=\! {(\bv\BS_0)}\herm \by\BS$ as in~\eqref{eq:rBS_k} upon setting $k\!=\!0$.

The DL signal received by UE $i$, $i \!\in\!\setwK$, of the IAB node is 
\begin{align*}
    \by\UE_i  &\!=\! \bH\herm_i \bx\IAB + \sum\nolimits_{k\in\setK} \sqrt{\eta\UE_k} \bH\herm_{i,k} \bbf\UE_k \bx\UE_k + \bn\UE_i \nonumber \\
    &\!=\! \bH\herm_i \!\!\!\! \sum\limits_{\ell \in \setwK\cup\{0\}} \!\!\!\! \sqrt{\eta\IAB_{\ell}} \bbf\IAB_{\ell} x\IAB_{\ell} \!\!+\!\! \sum\limits_{k \in \setK} \!\sqrt{\eta\UE_k} \bH\herm_{i,k} \bbf\UE_k \bx\UE_k \!+\! \bn\UE_i.    
\end{align*}
Then, UE $i$, $i \!\in\!\setwK$, linearly combines $\by\UE_i$ to decode $x\IAB_i$ as $r\UE_i \!=\! {(\bv\UE_i)}\herm\by\UE_i$ and obtain
\begin{align*}
    r\UE_i \!=\!\! \sum\limits_{\ell\in\setwK\cup\{0\}} \!\!\!\sqrt{\eta\IAB_{\ell}} \bg_{i,i,\ell}^{\mathsf{UE\text{-}IAB}} x\IAB_{\ell}  \!+\! \sum\limits_{k\in\setK} \sqrt{\eta\UE_k} \Tilde{\bg}_{i,k}^{\mathsf{UE\text{-}UE}} x\UE_k \!+\! \Tilde{\bn}\UE_i,  
\end{align*}
where $\Tilde{\bn}\UE_i \!=\! {(\bv\UE_i)}\herm\bn\UE_i$.

\section{Performance Analysis} \label{sec:performance-analysis}

\subsection{Achievable SE at the gNB}
The UL SE at the gNB over the access link, with respect to the UE $k$ transmitted data symbol $x\UE_k$, $k \!\in\! \setK$, and under the assumption of perfect channel state information (CSI) knowledge at the gNB, is given by
\begin{align}
\label{eq:SE:uplink:BS:UE}
\SEuBS_k = \EX{\log_2 (1+\SINRuBS_k)} \, , \text{[bit/s/Hz]},  
\end{align} 
where the effective SINR is
\begin{align}
\label{eq:SINR:uplink:BS:UE}
\SINRuBS_k \!=\! \frac{\eta\UE_k \sabs{\hat{\bg}_{k,k,k}^{\mathsf{gNB\text{-}UE}}}}{\INTuBS_k\!+ \eta\IAB_{0} \sabs{\hat{\bg}_{k,0,0}^{\mathsf{gNB\text{-}IAB}}} \!+\!\sigma^2_{\mathsf{gNB}}\snorm{\bv\BS_k}}\, ,
\end{align}
and with the multiuser interference being equal to
\begin{equation}
    \INTuBS_k \!=\! \sum\limits_{i \in \setwK} \eta\IAB_{i} \sabs{\hat{\bg}_{k,0,i}^{\mathsf{gNB\text{-}IAB}}}\!\!+\!\!\sum\limits_{j \in\setK\setminus\{k\}} \!\eta\UE_j \sabs{\hat{\bg}_{k,j,j}^{\mathsf{gNB\text{-}UE}}}.
\end{equation}
The UL SE at the gNB over the backhaul link, with respect to the IAB data symbol $x\IAB_0$, and under the assumption of perfect CSI knowledge at the gNB, is given by
\begin{align}
\label{eq:SE:uplink:BS:IAB}
\SEuBS_0 = \EX{\log_2 (1+\SINRuBS_0)} \, , \text{ [bit/s/Hz]},  
\end{align} 
where the effective SINR is
\begin{align}
\label{eq:SINR:uplink:BS:IAB}
\SINRuBS_0 = \frac{\eta\IAB_0 \sabs{\hat{\bg}_{0,0,0}^{\mathsf{gNB\text{-}IAB}}}}{\INTuBS_0 \!+\! \sigma^2_{\mathsf{gNB}}\lVert\bv\BS_0\rVert^2}\, ,
\end{align}
and with the interference contribution being equal to
\begin{equation}
    \INTuBS_0 \!=\! \sum\limits_{i \in \setwK} \eta\IAB_{i} \sabs{\hat{\bg}_{0,0,i}^{\mathsf{gNB\text{-}IAB}}} \!+\! \sum\limits_{j\in\setK} \eta\UE_j \sabs{\hat{\bg}_{0,j,j}^{\mathsf{gNB\text{-}UE}}}\,. 
\end{equation}

\subsection{Achievable SE at the UE}
The achievable DL SE at UE $k$, $k\!\in\!\setK$, served by the gNB, under the assumption of perfect CSI knowledge at the UE, is
\begin{align}
\label{eq:SE:downlink:BS}
\SEdBS_k = \EX{\log_2 (1+\SINRdBS_k)} \, , \text{ [bit/s/Hz]},  
\end{align} 
where the effective SINR is
\begin{align}
\label{eq:SINR:downlink:BS}
\SINRdBS_k = \frac{\eta\BS_k \sabs{{\bg}_{k,k,k}^{\mathsf{UE\text{-}gNB}}}}{\INTdBS_k + \sigma^2_k\snorm{\bv\UE_k}}\, ,
\end{align}
and with the interference contribution being equal to
\begin{equation}
    \INTdBS_k \!=\! \sum\limits_{\substack{j \in\setK\cup\{0\} \\ j\neq k}} \eta\BS_j \sabs{{\bg}_{k,k,j}^{\mathsf{UE\text{-}gNB}}} \!+\! \sum\limits_{i\in\setwK} \eta\UE_i \sabs{\Tilde{\bg}_{k,i}^{\mathsf{UE\text{-}UE}} }\,.
\end{equation}
The achievable DL SE at UE $i$, $i\!\in\!\setwK$, served by the IAB node, under the assumption of perfect CSI knowledge at the UE, is
\begin{align}
\label{eq:SE:downlink:IAB}
\SEdIAB_i = \EX{\log_2 (1+\SINRdIAB_i)} \, , \text{ [bit/s/Hz]},  
\end{align} 
where the effective SINR is
\begin{align}
\label{eq:SINR:downlink:IAB}
\SINRdIAB_i= \frac{\eta\IAB_i \sabs{{\bg}_{i,i,i}^{\mathsf{UE\text{-}IAB}}}}{\INTdIAB_i \!+\!\eta\IAB_{0} \sabs{{\bg}_{i,i,0}^{\mathsf{UE\text{-}IAB}}} \!+\! \sigma^2_i\snorm{\bv\UE_i}}\, ,
\end{align}
and with the multiuser interference being equal to
\begin{equation}
    \INTdIAB_i \!=\! \sum\limits_{k\in\setK}\eta\UE_k \sabs{\Tilde{\bg}_{i,k}^{\mathsf{UE\text{-}UE}}} \!\!+\!\! \sum\limits_{\ell\in\setwK\setminus\{i\}} \eta\IAB_{\ell} \sabs{{\bg}_{i,i,\ell}^{\mathsf{UE\text{-}IAB}}}\,.
\end{equation}
\begin{remark}
$\SEdIAB_0$, obtained from~\eqref{eq:SE:downlink:IAB} upon setting $i\!=\!0$, gives the achievable DL SE at the gNB with respect to the data symbol $x\IAB_0$ transmitted by the IAB node, namely the achievable SE over the UL backhaul link. Indeed, the gNB can be thought as an IAB node's UE in the UL operation. Hence, it holds that $\SEdIAB_0$ coincides with $\SEuBS_0$ given by~\eqref{eq:SE:uplink:BS:IAB}.
\end{remark}

\subsection{Achievable SE at the IAB} \label{subsec:SE-IAB}
The UL SE at the IAB node over the backhaul link, with respect to the symbol $x\BS_0$ transmitted by the gNB, and under the assumption of perfect CSI knowledge at the IAB node, is
\begin{align}
\label{eq:SE:uplink:IAB}
\SEuIAB_0 = \EX{\log_2 (1+\SINRuIAB_0)}\,, \text{ [bit/s/Hz]},  
\end{align} 
where the effective SINR is
\begin{align}
\label{eq:SINR:uplink:IAB}
\SINRuIAB_0 = \frac{\eta\BS_0 \sabs{\bg_{0,0,0}^{\mathsf{IAB\text{-}gNB}}}}{ \INTuIAB_0\!+\! \sigma^2_{\mathsf{IAB}}\snorm{\bv\IAB_0}}\,,
\end{align}
and with the interference contribution being equal to
\begin{equation}
    \INTuIAB_0 \!=\! \sum\limits_{k \in \setK} \eta\BS_k \sabs{\bg_{0,0,k}^{\mathsf{IAB\text{-}gNB}}} \!\!+\!\!\sum\limits_{i \in \setwK} \eta\UE_i \sabs{\hat{\bg}_{0,i,i}^{\mathsf{IAB\text{-}UE}}}\,.
\end{equation}
The UL SE at the IAB node over the access link, with respect to the UE $i$ transmitted data symbol, $x\UE_i$, $i\!\in\!\setwK$, and under the assumption of perfect CSI knowledge at the IAB node, is
\begin{align}
\label{eq:SE:uplink:IAB:UE}
\SEuIAB_i = \EX{\log_2 (1+\SINRuIAB_i)} \, , \text{ [bit/s/Hz]},  
\end{align} 
where the effective SINR is
\begin{align}
\label{eq:SINR:uplink:IAB:UE}
\SINRuIAB_i = \frac{\eta\UE_i \sabs{\hat{\bg}_{i,i,i}^{\mathsf{IAB\text{-}UE}}}}{ \INTuIAB_i + \sigma^2_{\mathsf{IAB}}\snorm{\bv\IAB_i}}\,,
\end{align}
and with the interference contribution being equal to
\begin{equation}
    \INTuIAB_i \!=\!\!\! \sum\limits_{\ell\in\setwK\setminus\{i\}} \!\!\!\eta\UE_{\ell} \sabs{\hat{\bg}_{i,\ell,\ell}^{\mathsf{IAB\text{-}UE}}} \!+\!\!\! \sum\limits_{k \in\setK\cup\{0\}} \!\!\! \eta\BS_k \sabs{\bg_{i,0,k}^{\mathsf{IAB\text{-}gNB}}}.
\end{equation}
\begin{remark}
$\SEdBS_0$, obtained from~\eqref{eq:SE:downlink:BS} upon setting $k\!=\!0$, gives the achievable DL SE at the IAB node with respect to the data symbol $x\BS_0$ transmitted by the gNB, namely the achievable SE over the DL backhaul link. Indeed, the IAB node can be thought as a gNB's UE in the DL operation. It holds that $\SEdBS_0$ coincides with $\SEuIAB_0$ given by~\eqref{eq:SE:uplink:IAB}.
\end{remark}

\section{Power Allocation and Beamforming} \label{sec:centralized_PA}
In this section, we propose and globally solve an optimization problem to properly determine the transmit powers at the gNB, IAB node, and UEs, in a centralized fashion.
Hence, this optimization problem couples UL and DL operations and is carried out at the gNB, after collecting CSI from the IAB node and the UEs.
Let us define the collective transmit power vectors for gNB, IAB node and UEs as $\boldsymbol{\eta}\BS = [\eta\BS_0, \eta\BS_1, \ldots, \eta\BS_K]\trans$, $\boldsymbol{\eta}\UE = [\eta\UE_1,\ldots,\eta\UE_{K\!+\!\wK}]\trans$, and $\boldsymbol{\eta}\IAB = [\eta\IAB_0, \eta\IAB_{K\!+\!1}, \ldots, \eta\IAB_{K\!+\!\wK}]\trans$, respectively. Then, the optimization problem for power allocation is generalized as
\begin{subequations} \label{prob:centralized:MR:original}
\begin{align}	
  \mathop {\text{maximize}}\limits_{\boldsymbol{\eta}\BS,~\boldsymbol{\eta}\IAB,~\boldsymbol{\eta}\UE} & \quad t\uBS_{\setK} + t\dBS_{\setK} + t\uIAB_{\setwK} + t\dIAB_{\setwK}  
  \label{prob:centralized:MR:obj} \\[-1ex] 	
  \text{s.t.} &\quad \sum\nolimits_{i \in \setwK} \SEdIAB_i \leq  \SEuIAB_0 \, ,
                \label{prob:centralized:MR:c1} \\
              &\quad \sum\nolimits_{i \in \setwK} \SEuIAB_i \leq  \SEdIAB_0 \, ,
                \label{prob:centralized:MR:c2} \\
              &\quad \eta\BS_0+\sum\nolimits_{k \in \setK}\eta\BS_k \leq \pMaxBS \, ,
                \label{prob:centralized:MR:c3} \\ 
              &\quad \eta\IAB_0+\sum\nolimits_{i \in \setwK}\eta\IAB_i \leq \pMaxIAB \, ,
                \label{prob:centralized:MR:c4} \\
              &\quad \eta\UE_k \leq \pMaxUE,~k=1,\ldots,K+\wK \, ,
                \label{prob:centralized:MR:c5}
\end{align}
\end{subequations}
where the objective consists of maximizing a utility involving either the SE or the SINR.
Specifically, we consider two power allocation (PA) strategies: $(i)$
Max-min fairness PA where $ t^{\mathsf{x,y}}_{\mathcal{X}} \!=\! \min\nolimits_{k \in \mathcal{X}} \, \SINRxy_k$, and $(ii)$ Max-sum SE PA where $t^{\mathsf{x,y}}_{\mathcal{X}} \!=\! \sum\nolimits_{k \in \mathcal{X}} \, \SExy_k$, with $\mathsf{x} \!=\! \{ \mathsf{d}, \, \mathsf{u} \}$, $\mathsf{y} \!=\! \{ \mathsf{gNB}, \, \mathsf{IAB} \}$, $\mathcal{X} \!=\! \{\setK,\, \setwK\}$. The peculiarity of Problem~\eqref{prob:centralized:MR:original} lies in the coupling between UL and DL operation set by the objective and in the SE constraints enforced by the gNB-to-IAB node backhaul link---the sum SE the IAB delivers to its UEs is constrained by the DL SE of the backhaul link \rev{(i.e., \eqref{prob:centralized:MR:c1})}, as well as the sum SE of the IAB uplink is constrained by the UL SE of the backhaul link    \rev{(i.e., \eqref{prob:centralized:MR:c2})}. Problem~\eqref{prob:centralized:MR:original} is generally non-convex due to the non-convexity of the objective function and the SE constraints~\refs{prob:centralized:MR:c1}{prob:centralized:MR:c2} by the SINR structures. Next, we present an approach to convexify the problem, enabling to achieve a global optimum for~\eqref{prob:centralized:MR:original} 
under the assumption of combining and precoding vectors that are independent of the transmit powers. In this regard, we assume maximum-ratio combining (MRC) and \textit{singular value decomposition} (SVD)-based precoding. We let the SVD of an arbitrary channel matrix $\bH_k \!\in\! \mC^{m\times n}$, $k \!=\! 0, \ldots, K \!+\! \wK$, be $\bH_k \!=\! \bU_k \boldsymbol{\Lambda}_k \bW_k\herm$, where $\bU_k \!=\! [\bu^{(k)}_1, \ldots, \bu^{(k)}_m]$ and $\bW_k \!=\! [\bw^{(k)}_1, \ldots, \bw^{(k)}_n]$ are $m \times r_k$ and $n \times r_k$, respectively, unitary matrices that contain the left and right singular vectors, with $r_k\!=\!\text{rank}(\bH_k)$. Besides, $\boldsymbol{\Lambda}_k$ is a $r_k \times r_k$ diagonal matrix with the singular values on the diagonal in a descending order. The precoding vectors are then set as $\bbf\BS_k \!=\! \bu^{(k)}_1$, $k\!\in\!\setK\cup\{0\}$, $\bbf\UE_k \!=\! \bw^{(k)}_1$, $k\!\in\!\setK \cup \setwK$, $\bbf\IAB_0 \!=\! \bw^{(0)}_1$, and $\bbf\IAB_k \!=\! \bu^{(k)}_1$, $k \!\in\! \setwK$, where $\bu^{(k)}_1$ ($\bv^{(k)}_1$) is the left (right) singular vector corresponding to the largest singular value of $\bH_k$. This choice meets the unitary norm requirements for the precoders set in~\Secref{Sec:SystemModel}. While, the MRC vectors are given~by $\bv\IAB_0 \!=\! \bH\herm_0\bbf\BS_0$, $\bv\IAB_i \!=\! \bH_i\bbf\UE_i$, $i \!\in\!\setwK$, $\bv\BS_k \!=\! \bH_k\bbf\UE_k$, $k \!\in\!\setK$, $\bv\BS_0 \!=\! \bH_0\bbf\IAB_0$, $\bv\UE_k \!=\! \bH\herm_k\bbf\BS_k$, $k \!\in\!\setK$, and $\bv\UE_i \!=\! \bH\herm_i\bbf\IAB_i$, $i \!\in\!\setwK$.

\subsection{Max-Min Fairness Power Allocation}
\label{subsec:max-min-PA}
In the max-min fairness PA problem $t^{\mathsf{x,y}}_{\mathcal{X}} \!=\! \min\nolimits_{k \in \mathcal{X}} \SINRxy_k$. Let us define the auxiliary vector $\bz \!=\! [z_1 \, z_2, \ldots, z_N]\trans \!\in\! \mathbb{R}_{+}^{N \times 1}$, with $N\! =\! 4$, and the vector $\boldsymbol{\mathsf{SINR}}_k \in \mathbb{R}_{+}^{4 \times 1}$ as
\begin{align*}
    \boldsymbol{\mathsf{SINR}}_k \!=\! \left[\mathsf{SINR}\uBS_{k,\,\setK}, \, \mathsf{SINR}\dBS_{k,\,\setK}, \, \mathsf{SINR}\uIAB_{k,\,\setwK}, \, \mathsf{SINR}\dIAB_{k,\,\setwK} \right]\trans, 
\end{align*}
where $\SINRxy_{k,\,\mathcal{X}} \!=\! \SINRxy_k$, with $k \!\in\! \mathcal{X} \!=\! \{\setK,\, \setwK\}$, $\mathsf{x} \!=\! \{ \mathsf{d}, \, \mathsf{u} \}$, and $\mathsf{y} \!=\! \{ \mathsf{gNB}, \, \mathsf{IAB} \}$, such that 
\begin{align} \label{eq:SINR_upper}
    \bz \preceq \boldsymbol{\mathsf{SINR}}_k, \, \forall k,~k \in \{\setK,\, \setwK\} \, , 
\end{align}
with $\preceq$ denoting the element-wise inequality. Problem~\eqref{prob:centralized:MR:original} can be reformulated as
\begin{subequations} \label{prob:max-min:MR:reformulation}
\begin{align}	
  \mathop {\text{maximize}}\limits_{\substack{\boldsymbol{\eta}\BS,~\boldsymbol{\eta}\IAB, \\ \boldsymbol{\eta}\UE,~\bz}} & \; \Big( \, \prod\nolimits_{n=1}^N z_n \Big)^{1/N} 
  \label{prob:centralized:MR:obj_r} \\[-1ex] 	
  \text{s.t.} &\quad \eqref{prob:centralized:MR:c1}, \eqref{prob:centralized:MR:c2}, \eqref{prob:centralized:MR:c3}, \eqref{prob:centralized:MR:c4}, \eqref{prob:centralized:MR:c5}\nonumber,\\
  &\quad \bz \preceq \boldsymbol{\mathsf{SINR}}_k, \, \forall k,~k \in \{\setK,\, \setwK\}
  \label{prob:centralized:MR:c6},
\end{align}
\end{subequations}
which follows from the inequality of arithmetic and geometric means, with $\{z_n\}^N_{n=1}$ being non-negative real numbers, 
\begin{align}\label{eq:geometric_mean}
    \frac{1}{N} \sum\limits_{n = 1}^{N} z_n \geq \Big( \, \prod\nolimits_{n=1}^N z_n\Big)^{1/N}.
\end{align}
Hence, Problem~\eqref{prob:max-min:MR:reformulation} involves the maximization of the monomial in \eqref{prob:centralized:MR:obj_r}, constrained to the power budgets in \eqref{prob:centralized:MR:c3}, \eqref{prob:centralized:MR:c4} and \eqref{prob:centralized:MR:c5}. The SE constraints \eqref{prob:centralized:MR:c1}-\eqref{prob:centralized:MR:c2} need a mathematical manipulation in order to recast the optimization problem into a generalized geometric program. First, by exploiting the logarithm property, \eqref{prob:centralized:MR:c1} and \eqref{prob:centralized:MR:c2} are rewritten, respectively,~as
\begin{align}
    \frac{\prod_{i \in \setwK}(1+\SINRdIAB_i)}{1+\SINRuIAB_0} &\leq 1\,, \label{eq:SE1} \\
    \frac{\prod_{i \in \setwK}(1+ \SINRuIAB_i)}{1+ \SINRdIAB_0} &\leq 1\,. \label{eq:SE2}
\end{align}
Second, we define the auxiliary variables
\begin{align}\label{eq:auxiliary:varrho}
\boldsymbol{\varrho}^{\mathsf{x}} &= \left[\varrho^{\mathsf{x}}_{K+1}, \, \varrho^{\mathsf{x}}_{K+2}, \ldots, \varrho^{\mathsf{x}}_{K+\wK}\right]\trans \in \mR_{+}^{\wK \times 1},   
\end{align}
and $z_0^{\mathsf{x}}$, with $\mathsf{x}\!=\!\{ \mathsf{d}, \, \mathsf{u} \}$, such that
\begin{numcases}{}
&$\!\!\!\varrho^{\mathsf{x}}_{i} \geq \SINRsIAB_i,\,\forall i \in \setwK,\, \mathsf{x}\!=\!\{ \mathsf{d}, \, \mathsf{u} \} $, \label{eq:SINR_geq} \\
&$\!\!\!z_0^{\mathsf{x}} \leq \SINRsIAB_0, \, \mathsf{x}\!=\!\{ \mathsf{d}, \, \mathsf{u} \}$, \label{eq:SINR_lower1}
\end{numcases} 
Hence, the SINR lower bound constraints \eqref{eq:SINR_lower1} are cast as geometric programming by taking the inverse of both sides \cite{boyd2007tutorial}. In contrast, the constraints \eqref{eq:SINR_geq} can be written as
\begin{align}
    \label{eq:SINR-constraint-reformulation}
    \frac{\varrho^{\mathsf{x}}_{i}}{\mathsf{num}\sIAB_i} \geq \frac{1}{\sum\nolimits^{M}_{m=1} \mathsf{den}\sIAB_{i,m}} \, , \, \forall i \in \setwK,\, \mathsf{x}\!=\!\{ \mathsf{d}, \, \mathsf{u} \}, 
\end{align}
where the term $\mathsf{num}\sIAB_i$ is the monomial at the numerator of $\SINRsIAB_i$, with $\mathsf{x}\!=\!\{ \mathsf{d}, \, \mathsf{u} \}$, in~\eqref{eq:SINR:uplink:IAB} or~\eqref{eq:SINR:downlink:IAB}, while $\mathsf{den}\sIAB_{i,m}$ is the $m$-th monomial at the denominator, and $M \!=\! K \!+\!\wK\!+\!1$.
By using~\eqref{eq:geometric_mean}, \eqref{eq:SINR-constraint-reformulation} can be relaxed (and convexified) to 
\begin{align}\label{eq:constraint:SINR:dIAB}
  \frac{M \, \varrho^{\mathsf{x}}_{i} \left(\prod\nolimits^M_{m=1} \mathsf{den}\sIAB_{i,m}\right)^{1/M}}{\mathsf{num}\sIAB_i} \geq 1 \, , \, \forall i \in \setwK,\,\mathsf{x}\!=\!\{ \mathsf{d}, \, \mathsf{u} \}.
\end{align}
Combining all the above results, constraints \eqref{eq:SE1}-\eqref{eq:SE2} read as
\begin{align}
   \prod\nolimits_{i \in \setwK}(1 \!+\! \varrho^{\mathsf{d}}_{i}) &\!\leq\! 1 \!+\! z_0^{\mathsf{u}}, \label{eq:relaxA}\\
   \prod\nolimits_{i \in \setwK}(1 \!+\! \varrho^{\mathsf{u}}_{i}) &\!\leq\! 1 \!+\! z_0^{\mathsf{d}}, \, \label{eq:relaxB}
\end{align}
respectively, and are cast as geometric constraints.
Finally, problem \eqref{prob:max-min:MR:reformulation} is reformulated as
\begin{subequations} \label{prob:max-min:MR:final}
\begin{align}	
  \mathop {\text{maximize}}\limits_{\substack{\boldsymbol{\eta}\BS,\, \boldsymbol{\eta}\IAB,\, \boldsymbol{\eta}\UE \\ \bz, \, \boldsymbol{\varrho}^{\mathsf{d}}, \, \boldsymbol{\varrho}^{\mathsf{u}}, \, z_0^{\mathsf{u}}, \, z_0^{\mathsf{d}}}} &\; \left(\, \prod\limits_{n=1}^N z_n \right)^ {1/N} \label{prob:centralized:MR:obj_rr} \\ 	
  \text{s.t.} &\;\; \eqref{prob:centralized:MR:c3}, \eqref{prob:centralized:MR:c4}, \eqref{prob:centralized:MR:c5},\eqref{eq:SINR_upper}\nonumber,\\
  &\;\; \eqref{eq:SINR_lower1}, \eqref{eq:constraint:SINR:dIAB}, \eqref{eq:relaxA}, \eqref{eq:relaxB},\nonumber
\end{align}
\end{subequations}
which constitutes a geometric program. It admits global optimum and can be efficiently solved via interior-point methods by using common solvers.

\subsection{Max-Sum SE Power Allocation}
The sum-SE maximization entails solving Problem~\eqref{prob:centralized:MR:original} with $t^{\mathsf{x,y}}_{\mathcal{X}} \!=\! \sum\nolimits_{k \in \mathcal{X}} \, \SExy_k$. Compared to the approach detailed in~\Secref{subsec:max-min-PA}, we need to reformulate~\eqref{eq:SINR_upper} as $\bz \!\preceq\! \bs$, where
\begin{equation}
    \bs \!=\! \left[\sum_{k \in \setK} s\uBS_{k,\setK}, \, \sum_{k \in \setK} s\dBS_{k,\setK}, \, \sum_{k \in \setwK} s\uIAB_{k,\setwK}, \, \sum_{k \in \setwK} s\dIAB_{k,\setwK} \right]\trans,
\end{equation}
and $s^{\mathsf{x,y}}_{k,\mathcal{X}}$ being a set of auxiliary variables with $\mathsf{x} \!=\! \{ \mathsf{d}, \, \mathsf{u} \}$, $\mathsf{y} \!=\! \{ \mathsf{gNB}, \, \mathsf{IAB} \}$ and $\mathcal{X} \!=\! \{\setK,\, \setwK\}$ necessary to handle the SE expressions and subject to the following constraints:
\begin{align}
    s^{\mathsf{x,y}}_{k,\mathcal{X}} \leq  \log_2(1+\mathsf{SINR}^{\mathsf{x,y}}_k)\,, \forall k \in \mathcal{X}.
\end{align}
The latter can be rewritten as 
\begin{align}
\exp{\left\{{\ln(2) s^{\mathsf{x,y}}_{k,\mathcal{X}}}\right\}} \leq 1+\mathsf{SINR}^{\mathsf{x,y}}_k\,, \forall k \in \mathcal{X},
\end{align}
which, by using the approximation in~\cite{boyd2007tutorial} for the left-hand side, can be cast to the following geometric programming form 
\begin{align}
\label{eq:SE-constraint-reformulation}
        {\left(1 \!+\! \frac{\ln(2)}{\varepsilon} s^{\mathsf{x,y}}_{k,\mathcal{X}}\right)}^{\displaystyle\!\varepsilon} \!\leq 1+ \mathsf{SINR}^{\mathsf{x,y}}_k\,, \forall k \in \mathcal{X},
\end{align} 
with $\varepsilon$ being a sufficiently large positive even constant.
Hence, the max-sum SE PA problem is formulated as~\eqref{prob:max-min:MR:final}, but replacing~\eqref{eq:SINR_upper} with the element-wise constraint $\bz \!\preceq\! \bs$, and adding the set of constraints~\eqref{eq:SE-constraint-reformulation}.

\section{Numerical Results} \label{sec:simulation}

\begin{figure}[t]
    \centering
    \includegraphics[width=0.98\columnwidth]{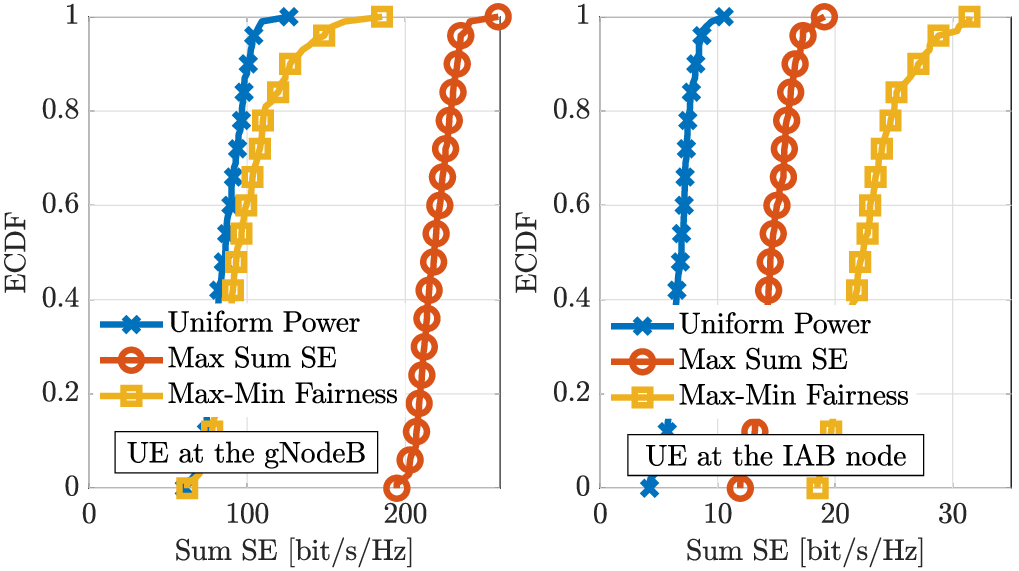}
    \vspace*{-3mm}
    \caption{ECDF of Sum SE with $K \!=\! 12$ UEs in gNB area and $\wK \!=\! 1$ UE in IAB node area, comparing different power allocation schemes.}
    \vspace*{-2mm}
    \label{fig:ecdf_k1}
\end{figure}

\begin{figure}[t]
    \centering
    \includegraphics[width=0.98\columnwidth]{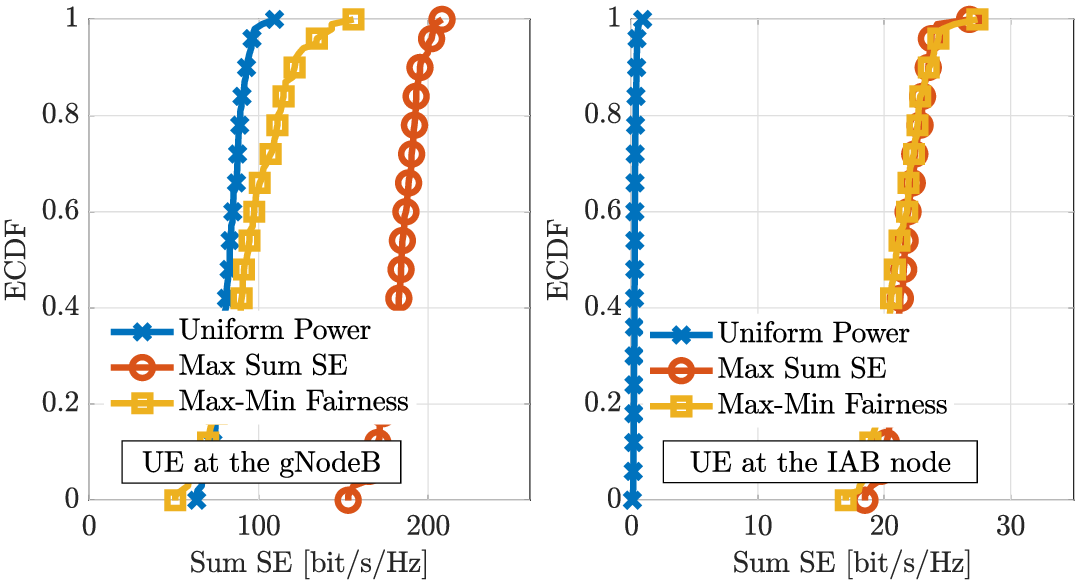}
    \vspace*{-3mm}
    \caption{ECDF of Sum SE with $K \!=\! 12$ UEs in gNB area and $\wK \!=\! 10$ UEs in IAB node are, comparing different power allocation schemes.}
    \vspace*{-4mm}
    \label{fig:ecdf_k10}
\end{figure}

\begin{figure}[t]
    \centering
    \includegraphics[width=0.99\columnwidth]{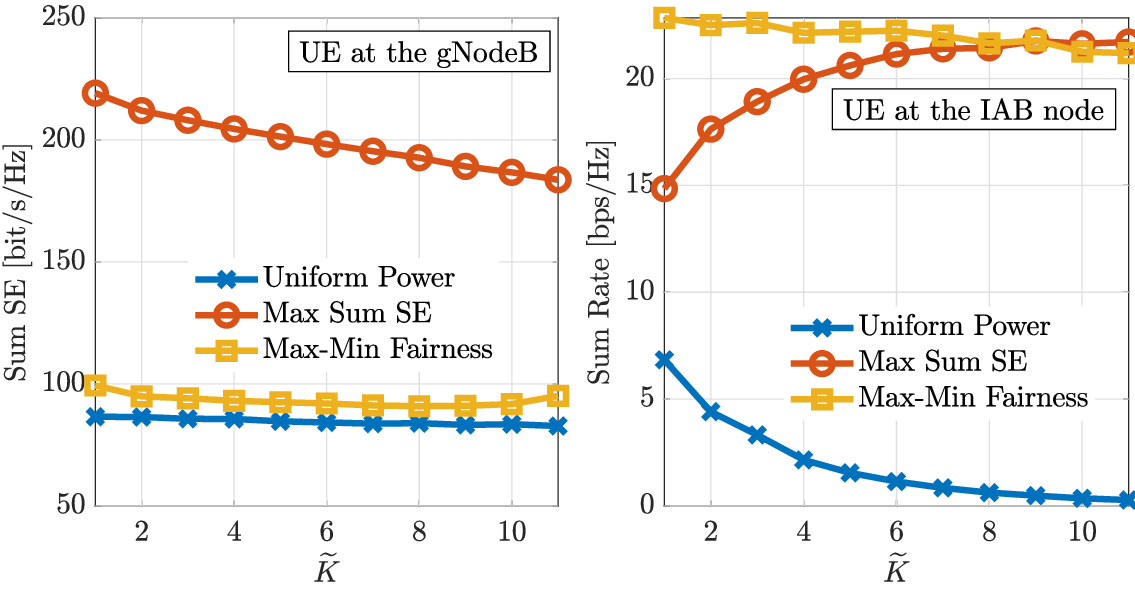}
    \vspace*{-8mm}
    \caption{Average Sum SE versus $\wK$ assuming $K \!=\! 12$ UEs in gNB area, comparing different power allocation scheme.}
    \vspace*{-5mm}
    \label{fig:avg_sum_rate}
\end{figure}

This section provides a detailed analysis of system performance by comparing various power allocation strategies. We focus on the sum SE by aggregating UL and DL SE of all the UEs located in the IAB node and the gNB area. Unless otherwise specified, the simulation parameters outlined in Table \ref{tab:simulation_parameters} serve as the basis for our evaluations.
\begin{table}[!t]
    \centering
    \caption{Simulation Parameters}
    \begin{tabular}{|c|c|}
    \hline
    \textbf{Simulation Parameter} & \textbf{Value} \\
    \hline \hline
    Carrier Frequency & 30\ GHz \\
    Channel Model & 3GPP TR 38.901 (UMI)~\cite{3GPPTR38901} \\
    Noise power density & -173\ dBm/Hz \\
    No. of antennas: $N\BS$, $N\IAB$, $N\UE$  & 16$\times$4, 8$\times$4, 4$\times$2 \\
    Max power: $\pMaxBS$, $\pMaxIAB$, $\pMaxUE$ & 43 dBm, 43 dBm, 23 dBm\\
    No. of gNB's UEs, $K$ & 12 \\
    Coverage radius: gNB, IAB & 100 m, 50 m \\
    UE distribution & Uniform\\
    Antenna type & 120-degree sector antenna\\
    \hline
    \end{tabular}
    \vspace*{-4mm}
    \label{tab:simulation_parameters}
\end{table}
The results depicted in Figs.~\ref{fig:ecdf_k1}, \ref{fig:ecdf_k10}, and \ref{fig:avg_sum_rate}, offer a comparative analysis of three power allocation schemes: uniform power allocation (serving as a benchmark), max-sum SE, and max-min fairness. Figs. \ref{fig:ecdf_k1} and \ref{fig:ecdf_k10}, show the empirical cumulative distribution function (ECDF) of the sum SE, differentiating UEs in the gNB area and those within the IAB node area for two scenarios: $\wK \!=\! 1$ (Fig. \ref{fig:ecdf_k1}) and $\wK \!=\! 10$ (Fig. \ref{fig:ecdf_k10}), where $\wK$ represents the number of UEs served by the IAB node. 
As illustrated in Figs. \ref{fig:ecdf_k1} and \ref{fig:ecdf_k10}, uniform power allocation consistently underperforms, highlighting the necessity for power optimization in IAB networks. The max-sum SE approach achieves the highest overall sum rate, which is particularly advantageous for UEs in the gNB area. Interestingly, when the number of UEs served by the IAB node is small, the max-min fairness scheme demonstrates superior performance for UEs within the IAB node area. However, as the number of UEs $\wK$ grows, interference increases, reducing the performance gap between the max-sum SE and the max-min fairness approaches, ultimately leading to similar results.
A critical observation from these figures is that the IAB backhaul link operates as a rank-1 channel, imposing significant performance limitations. This restriction is primarily due to the geometrical configuration of the scenario, where both the IAB node and the gNB are positioned in elevated locations, such as on buildings or dedicated towers. This clear line of sight minimizes the likelihood of multipath propagation, eventually restricting the backhaul capacity to a single spatial stream. Consequently, this bottleneck hinders the effectiveness of the IAB node in spatially multiplexing UEs.

Fig. \ref{fig:avg_sum_rate} shows the average sum SE that varies with $\wK$. The results show a decrease in the sum SE achieved by the max-sum SE approach as $\wK$ grows due to increased interference from UEs in the IAB node service area. The observations in Figs. \ref{fig:ecdf_k1} and \ref{fig:ecdf_k10} apply consistently here, reinforcing that the limited backhaul capacity fundamentally constrains performance.

\section{Conclusions}\label{sec:conclusion}

We examined the performance of an IAB node as a range extender for beyond-5G networks, focusing on the key challenges of effective power allocation and beamforming strategies, which are vital for maximizing UEs' sum SE.
We globally solved two power allocation optimization problems: max-sum SE and max-min fairness, to assess their effects on system performance. The results highlight the necessity of power optimization, particularly as the number of UEs served by the IAB node increases, showing how efficient power allocation enhances service quality in high-load scenarios.

Our findings revealed that while the max-sum SE consistently yields the best performance, max-min fairness is beneficial with fewer UEs in the IAB node service area, effectively managing interference. However, as UE numbers rise, these strategies' performance tends to converge. A limitation of the IAB node as a range extender stems from the bottleneck created by the rank-1 channel, arising from the high elevation of IAB nodes. This positioning reduces multipath opportunities and restricts backhaul capacity to a single spatial stream.

Future research could explore advanced beamforming techniques for improved interference mitigation, multi-hop IAB architectures, and multi-rank channel strategies to overcome rank-1 limitations, thereby enhancing multipath diversity and backhaul capacity.

\vspace*{-2mm}
\renewcommand{\baselinestretch}{.94}
\bibliographystyle{IEEEtran}
\bibliography{reference}

\end{document}